\documentclass[pdflatex,sn-mathphys-num]{sn-jnl}

\usepackage{orcidlink}

\usepackage{graphicx}%
\usepackage{multirow}%
\usepackage{amsmath,amssymb,amsfonts}%
\usepackage{amsthm}%
\usepackage{mathrsfs}%
\usepackage[title]{appendix}%
\usepackage{xcolor}%
\usepackage{textcomp}%
\usepackage{manyfoot}%
\usepackage{booktabs}%
\usepackage{algorithm}%
\usepackage{algorithmicx}%
\usepackage{algpseudocode}%
\usepackage{listings}%
\usepackage{aas_macros}
\usepackage{url}

\AtBeginShipoutNext{\AtBeginShipoutUpperLeft{%
  \put(\dimexpr\paperwidth-1cm\relax,-1.5cm){\makebox[0pt][r]{MIT-CTP/5848}}%
}}

\newcommand{\CMSSW}{CMSSW}
\newcommand{\ROOT}{ROOT}
\newcommand{\XROOTD}{XRootD}
\newcommand{\RDF}{RDataFrame}
\newcommand{\submit}{SubMIT}
\newcommand{\jupyterhub}{JupyterHub}
\newcommand{\coffea}{Coffea}
\newcommand{\nano}{NanoAOD}

\newcommand{\slurm}{Slurm}
\newcommand{\htc}{HTCondor}
\newcommand{\Ttwo}{Tier2}
\newcommand{\NVME}{NVMe}
\newcommand{\corrLib}{CorrectionLib}
\newcommand{\DaskGateway}{Dask Gateway}
\newcommand{\centoGbps}{100\,Gbps}
\newcommand{\ceph}{Ceph}

\theoremstyle{thmstyleone}%
\theoremstyle{thmstyletwo}%

\theoremstyle{thmstylethree}%

\begin{document}

\title[SubMIT]{SubMIT: A Physics Analysis Facility at MIT}

\author[1,2]{J.~Bendavid\orcidlink{0000-0002-7907-1789}}
\author[1]{M.~D'Alfonso\orcidlink{0000-0002-7409-7904}}
\author[1]{J.~Eysermans\orcidlink{0000-0001-6483-7123}}
\author[1]{C.~Freer\orcidlink{0000-0002-7967-4635}}
\author[1]{M.~Goncharov}
\author[1]{M.~Heine\orcidlink{0000-0002-4882-6712}}
\author[1]{L.~Lavezzo\orcidlink{0000-0002-1364-9920}}
\author[1]{M.~Moore\orcidlink{0000-0002-6424-0594}}
\author[1]{C.~Paus\orcidlink{0000-0002-6047-4211}}
\author[1]{X.~Shen\orcidlink{0000-0002-6196-823X}}
\author[1]{D.~Walter\orcidlink{0000-0001-8584-9705}}
\author[1]{Z.~Wang\orcidlink{0000-0002-3074-3767}}
\affil[1]{Massachusetts Institute of Technology (MIT), Cambridge, USA}
\affil[2]{CERN, Geneva, Switzerland}


\email{josh.bendavid@cern.ch}
\email{mariadlf@mit.edu}
\email{jaeyserm@mit.edu} 
\email{chadwfreer@gmail.com}
\email{maxi@mit.edu}
\email{mheine@mit.edu}
\email{lavezzo@mit.edu}
\email{mamoore@mit.edu}
\email{paus@mit.edu}
\email{xuejian@mit.edu}
\email{david\_w@mit.edu}
\email{wangzqe@mit.edu}


\abstract{The recently completed {\submit} platform is a small set of servers that provide interactive access to substantial data samples at high speeds, enabling sophisticated data analyses with very fast turnaround times. Additionally, it seamlessly integrates massive processing resources for large-scale tasks by connecting to a set of powerful batch processing systems. It serves as an ideal prototype for an Analysis Facility tailored to meet the demanding data and computational requirements anticipated during the High-Luminosity phase of the Large Hadron Collider. The key features that make this facility so powerful include highly optimized data access with a minimum of {\centoGbps} networking per server, a large managed NVMe storage system, and a substantial spinning-disk Ceph file system. The platform integrates a diverse set of high multicore CPU machines for tasks benefiting from the multithreading and GPU resources for example for neural network training. {\submit} also provides and supports a flexible environment for users to manage their own software needs for example by using containers. This article describes the facility, its users, and a few complementary, generic and real-life analyses that are used to benchmark its various capabilities.}




\maketitle

\section{Introduction}

The particle physics experiments at the High Luminosity Large Hadron Collider (HL-LHC) will generate an unprecedented volume of scientific data, reaching the multi-exabyte scale within the next decade. The computing model currently employed by the Large Hadron Collider (LHC), hosted at CERN, will not provide the required data processing power or storage capabilities, even when accounting for the anticipated hardware evolution during that time scale. Future big data experiments, such as the Deep Underground Neutrino Experiment (DUNE), the Large Synoptic Survey Telescope (LSST), the Square Kilometer Array (SKA), and the Electron-Ion Collider (EIC) face similar challenges.
To address these issues, collaborations must adopt new concepts and tools while developing analysis facilities that enable efficient and effective user analysis of these massive datasets.

This document outlines MIT's initiative to build a coherent and flexible demonstrator project of a CMS Analysis Facility (AF) called {\submit}, to investigate concepts for managing and analysing HL-LHC data. The {\submit} computing system was built in the context of the MIT Physics Department and supports research computing tasks for multiple high energy physics (HEP) experiments, including CMS, LHCb at LHC and for the feasibility study of the Future Circular Collider (FCC). It also provides computing resources for the entire physics department, including but not limited to the Laboratory of Nuclear Science (LNS), the Kavli Institute for Astrophysics and Space Research (MKI), the Center for Theoretical Physics (CTP), and the Center for Quantum Engineering (CQE). Given its large and diverse user base, {\submit} accommodates a wide range of workflows and is not exclusively tailored to HEP applications.

This note provides an in-depth examination of {\submit}'s design, deployment, configuration, optimization, and management.
Section~\ref{concepts} elaborates some of the foundational concepts and tools that have guided the system's design and development. Section~\ref{AF} introduces the essential features of an AF and various prototypes already implemented. The hardware and software configurations specific to {\submit} are detailed in Section~\ref{description}, followed by an overview of user engagement in Section~\ref{user}.  Performance evaluations and diverse use cases are documented in Section~\ref{analysis}, providing real-world applications and assessments of the system's capabilities. Finally, conclusions and future directions are presented in Section~\ref{conclusion}.

\section{Concepts and tools}
\label{concepts}
As the field of high energy physics has grown, so have the size and complexity of its datasets. In parallel with advancements in hardware and the development of new software techniques and algorithms, the concepts and tools used by analysers must evolve to optimally use the available resources.

Key developments include efficient data access, scalable computing strategies, machine learning techniques, and advanced workflow management systems that optimize resource allocation. Additionally, new approaches such as heterogeneous computing with specialized hardware for high-performance tasks, and intelligent caching strategies to minimize data movement, are becoming increasingly essential.

With these considerations in mind, this section discusses key factors that should be taken into account when designing a modern facility for efficient physics data analysis in the HL-LHC era.

\subsection{Distributed computing and software}
Since different physics events are statistically independent, parallelizing the workflow on different data chunks is a valid and very effective strategy to process the large physics datasets. From the very beginning of the LHC operations, the Worldwide LHC Computing Grid (WLCG)~\cite{WLCG} has played a crucial role in storing, distributing, and analysing data through its multi-tiered computing system. This approach takes advantage of computing power provided by different institutions that collaborate with CERN, coordinated through a unified computing grid.

For almost 20 years, the CMS Collaboration has used the {\CMSSW} software framework to successfully process both data and Monte Carlo workflows using multiple threads.
Given the modularity, {\CMSSW} has also been used for many different tasks including turning raw data into something physicists can use in an end-user analysis.
The CernVM-File System (CVMFS)~\cite{CVMFS} is a scalable, reliable and low-maintenance software distribution service that was developed to assist HEP collaborations to deploy software on the worldwide-distributed computing infrastructure used to run data processing applications.

Typical examples of analysis libraries adopted in HEP include \ROOT~\cite{ROOT}. This framework is mainly implemented in C++ but also offers Python bindings. The framework defines a common data structure and data layout to store HEP datasets, called TTree. Its layout is columnar on the disk, so that different columns can be treated independently. The \ROOT~I/O subsystem is able to read just a portion of a dataset, to minimize read requests to the file system.

\subsection{Data format compactification}
At each LHC experiment, data and simulated events undergo a standardized processing chain. To make large event samples more manageable, highly compressed datasets have been developed that contain only the necessary details on high-level physics objects for analysis: \nano~\cite{NANOAOD} at CMS and PHYSLITE~\cite{PHYSLITE} at ATLAS. These new event data formats enable faster and broader access and also minimize the need for experiment-specific software, thus enabling a data science approach to physics analyses.

On the other hand, some analyses that rely on rare or very detailed event information cannot be performed on these compact datasets. The experiments have to maintain the ability to access more detailed information. At CMS, this is addressed by the possibility of producing custom {\nano} where missing information can be added. This maintains the compact and standardized data format to simplify shared usage between different analyses. 

\subsection{Columnar analysis and services}
The end-user data analysis workflow continuously evolves to adapt to available hardware and software tools. In recent years, columnar and descriptive analysis frameworks, which leverage cloud technologies and heterogeneous resources, have made major advances. 

High-level interface tools for data analysis include {\RDF}~\cite{RDataFrame}.
{\RDF} is the interface for analysing a TTree in \ROOT, designed with parallelism in mind.
It can utilize all cores of a single machine via \ROOT's implicit multithreading interface and can scale by distributing computations across multiple nodes, dividing the analysis into tasks. It scales well to many cores, many nodes, and many histograms. 

New tools are being developed, including Python-based solutions like Uproot, Awkward-Array~\cite{Akward}, and \coffea~\cite{COFFEA}. Additional services like ServiceX~\cite{servicex} and Xcache~\cite{xcache} focus on providing users with high-throughput access to data. Services like the Dask Python library~\cite{dask} or Apache Spark~\cite{spark} provide users access to parallel computing through distributed clusters.

Both {\RDF} and {\coffea} provide a declarative solution to efficiently augment or filter data stored in the {\nano} file type, significantly speeding up throughput compared to traditional Python tools. These computations make any merging steps unnecessary, and allow one to directly produce final histograms in a single event loop.

\section{The Analysis Facility}
\label{AF}
To build a modern AF, analysis must shift from the conventional WLCG model -- with its high entry barriers and latency -- to a more interactive approach that speeds up data exploration. This reduces overhead, letting researchers focus on science rather than data handling and computing logistics.

A well-designed AF should facilitate access to software, storage, and computing resources. 
Emerging technologies must be continuously tested and evaluated for optimal integration with the existing WLCG infrastructure.
The HEP Software Foundation community released a white paper~\cite{Ciangottini:2024vtl} summarizing the current status and outlining the baseline R\&D needs for HL-LHC.

\subsection{Essential features}

Considering the complexities of modern computing, the principal components of an AF are its flexibility and ease of use. Essential features include: 
\begin{itemize}
    \item secure and straightforward access to the facility,
    \item interactive analysis on a single node with the ability to scale out to batch resources,
    \item opportunities for software customization, such as containerized solutions that can run at different facilities,
    \item efficient data access, facilitated by high-speed network connections to local storage systems and seamless retrieval of experimental data from larger computing facilities using standardized protocols,
    \item substantial computing power, including multicore CPU and GPU machines with sufficient memory, and
    \item comprehensive user support encompassing web pages, documentation, help desks, AI chat support, monitoring tools, meetings, and workshops between users and the computing support team.
\end{itemize}

\subsection{HEP prototypes}

A well established prototype of an interactive AF is SWAN (Service for Web-based ANalysis)~\cite{SWAN} at CERN. It is a web-based platform, built on Jupyter~\cite{Jupyter} notebooks, accessible to all CERN users, and provides immediate access to the CERN storage, software, and computing resources needed for analysis. It also allows access to external Spark clusters, multicore nodes, and machines equipped with accelerators, further enhancing computational capabilities. It is built on top of a local Kubernetes~\cite{kubernetes} cluster and integrates dedicated resources allocated via fair share through an \htc~scheduler~\cite{condor}.

Other prototypes for HEP experiments are being developed at various US institutions. These include Coffea-Casa~\cite{Nebraska} at the University of Lincoln, Nebraska; the Elastic Analysis Facility (EAF) at Fermilab~\cite{EAF} and Purdue~\cite{Purdue} for CMS. The ATLAS experiment is working on prototypes at SLAC, BNL, and the University of Chicago~\cite{USatlasAF}. These prototypes are all based on Dask and Jupyter notebook technologies built on top of Kubernetes clusters. More prototypes exist outside the US, for example, at DESY~\cite{DESY-NAF} and INFN~\cite{INFN}.

\section{Implementation of a full prototype: \submit}
\label{description}

\begin{figure}[t!]
    \centering
 \includegraphics[width=0.9\textwidth]{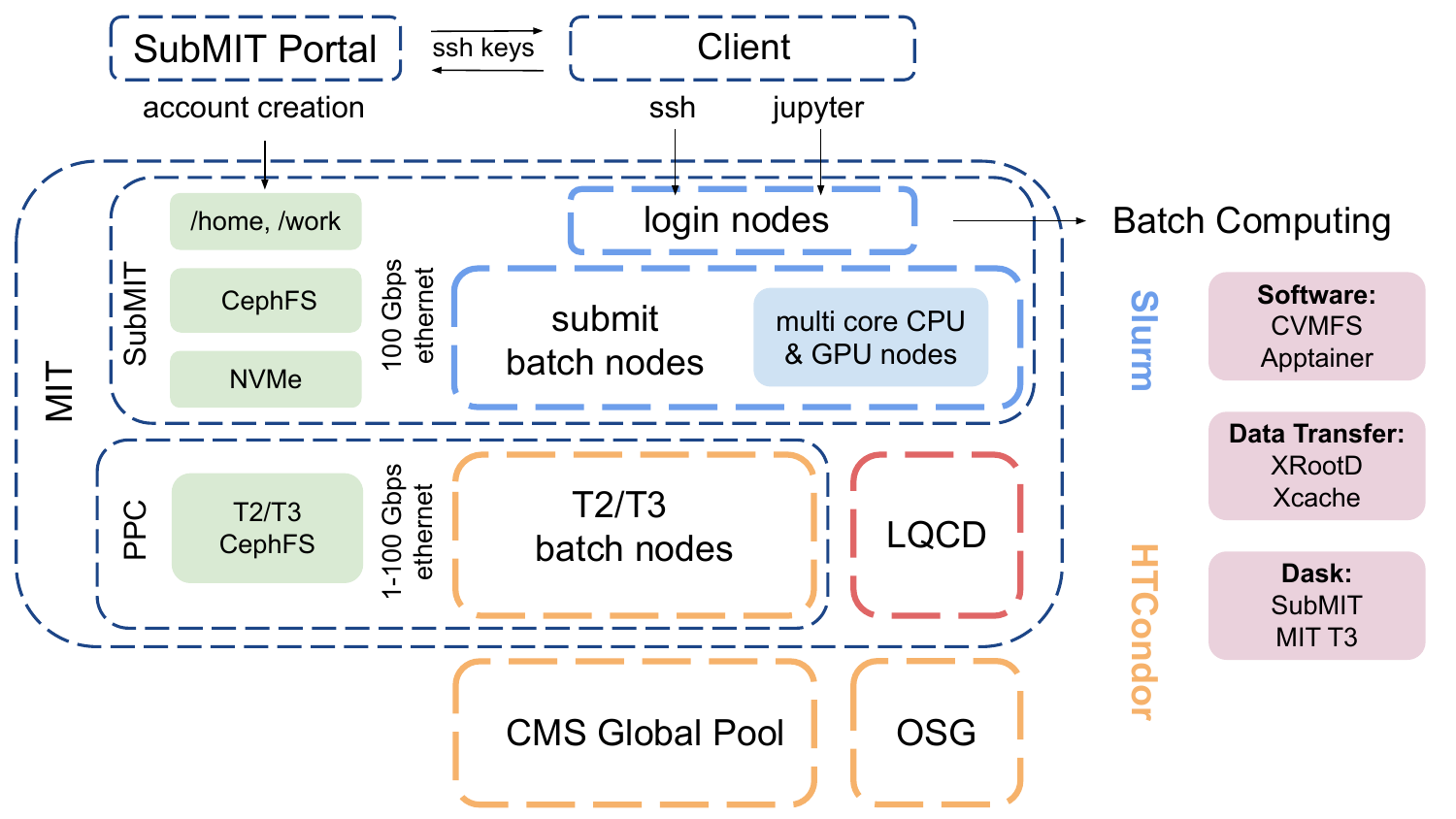}\\
    \caption{Schematic view of the {\submit} components and relations to other local and global clusters. Local SubMIT resources support batch computing via Slurm as highlighted in blue. Global clusters for batch computing via HTCondor highlighted in yellow are the CMS tier 2 (T2) and tier 3 centers (T3) at MIT, the CMS global computing pool, and the open science grid (OSG). The Lattice QCD (LQCD) cluster at MIT is also accessible. Different storage systems are colored in green.}
    \label{fig:AFlayout}
\end{figure}

The \submit~facility is an implementation of the AF concept at the MIT Department of Physics. Computing resources and storage areas are made available for interactive usage through a login pool, which can be seamlessly expanded to batch resources. This Section details the equipment and key functionalities as of 2025, with provisions for scalability to accommodate anticipated future enhancements. Currently, the services an AF should provide are not well defined within the community, and we expect that, between now and the HL-LHC era, the underlying technology will evolve as the community gains experience with new analysis methods. However, the primary objective is to abstract the infrastructure for the user. Our {\submit} prototype is actively validated by publishing CMS and other physics analyses, ensuring its continuous effectiveness in high-energy physics and other physics research. Additionally, it supports a broad range of projects within the department, including student classes. Figure~\ref{fig:AFlayout} summarizes the design of its various elements, which are further detailed in the upcoming subsections.

\subsection{Account creation}
Account creation for \submit~is streamlined for everyone with an MIT ID to accommodate undergraduate students who use the facility as an educational resource. The account is automatically created upon approval from the \submit~team, including the initialization of the user's home directory and storage areas. Individuals without an MIT ID can be sponsored for a guest account. At account creation, users are associated to a group to ensure the safely sharing input and output among different collaborations. Users are assigned an ID/GID through the MIT Lightweight Directory Access Protocol (LDAP) system, preventing potential clashes among MIT resources. 
Two-factor authentication is required for users accessing \submit~providing additional security.
Users can enable ssh login by uploading a public SSH key on the website.

\subsection{User interface}
The user has two main options to access {\submit}: either through SSH or through a web interface with the \jupyterhub~\cite{JupyterHub}~portal.
Both options enable interactive usage, full access to the storage system and available software, and allow users to dispatch their batch jobs to dedicated computing resources.

Eight interactive hosts are available for SSH logon for interactive usage and launching low-latency workflows. These machines, known as the \submit~login pool, provide a testing area for users to write, debug, and test their workflows before scaling them to batch systems. 

The \submit~system supports a custom \jupyterhub~instance. This instance is centrally maintained and secured behind the MIT's Shibboleth single sign-on (SSO) system. 
The central instance spawns individual Jupyter servers for individual users to any {\submit} node via \slurm~\cite{slurm}. Users have several options for the servers they want to spawn using \jupyterhub, whose resources and priorities can be customized based on demand via the {\slurm} job configurations. Once on the {\jupyterhub} servers, users can start personal notebooks, access all storage mounted on \submit, edit files, start Dask jobs, and execute any local code via a built-in terminal.

\subsection{Storage and data access}
Each user is initially allocated the following storage resources:
\begin{itemize}
    \item 5~GB redundant disk array as user’s home with backed-up storage notebooks and local software development (\verb|/home/submit/<user>|),
    \item 50~GB quota as user’s work for software installation (\verb|/work/submit/<user>|),
    \item 1~TB for user’s data to store large datasets (\verb|/ceph/submit/data/<user>|).
\end{itemize}
Additional space can be made available upon request and further include:
\begin{itemize}
    \item a server with 46~TB of \NVME~storage scratch disk space for fast access by groups, with room for future expansion (\verb|/scratch/submit/<user>|).
\end{itemize}
All {\submit} disks are accessible via a network of {\centoGbps} Ethernet to guarantee fast I/O. These spaces are visible from within every container which preserves the credential of the user. These spaces are also accessible for other users at the same AF or other sites, facilitating collaborative work among analysts pursuing the same analysis. The current mass storage system (\verb|/ceph/submit|) consists of 1.5\,PB of spinning disks in a CephFS distributed filesystem~\cite{ceph}. 

Data from the LHC experiments and their simulations are stored in mass storage solutions like \Ttwo/EOS and accessed via \XROOTD~\cite{XROOTD} protocol. The facility leverages local group disk storage from the MIT \Ttwo~(15 PB) with 2\,PB allocated for analysis purposes on a CephFS distributed filesystem. The access is typically through for LHC resources as well as group space on \verb|/ceph/submit| and the MIT \Ttwo. There is a perceived requirement to reduce local storage needs at analysis facilities by filtering and projecting input data and caching results and thus eliminate the need for manual bookkeeping and intermediate data storage by analysts. 
To enhance data access efficiency, {\submit} utilizes an XCache service located at the ESnet site in Cambridge (\verb|root://bost-cms-xcache01.es.net/|). This service provides cache space for frequently accessed data, leveraging high-performance {\NVME} storage devices to ensure rapid data retrieval.

\subsection{Computing resources}
Additional resources for medium-sized computational tasks apart from the login nodes are provided by the {\submit} system with approximately 2500 CPU cores.  In this way, users have the capability to scale out processing to execute distributed parallelization of computationally more intensive tasks. 
{\submit} systems is coupled with 2 GB of main memory per core. Specialized resources are available on demand, including 
\begin{itemize}
    \item two machines each equipped with two 96 core CPUs making 384 threads and 1.5\,TB of memory. One machine with two 192 core CPUs and 768 threads and 1.5\,TB of memory. These are designed for analysis tasks benefiting from multithreading over parallelization and
    \item twelve machines featuring accelerators, including 38 NVIDIA cards (2 RTX A6000, 12 V100s, 8~A30s, and 16 GeForce GTX 1080s), for machine learning development and the deployment of the \submit~chatbot. The various types of GPUs are available for batch or interactive use through {\slurm} scheduling.
\end{itemize}

\subsection{Batch computing}
Distributed computing, as implemented in the WLCG, has been a key strategy in HEP for efficiently processing large datasets through highly parallelized analyses of statistically independent events. Users can expand to external resources at MIT \Ttwo~(25k cores) and Tier-3 (700 cores). Other external connections require a grid certificate and include sites in different geographical locations, as defined in the CMS Global Pool (CMS users only) and the Open Science Grid (OSG) Global Pool. Additional connections to other MIT resources including the Lattice QCD (LQCD) clusters~\cite{lqcd} are enabled. On \submit, the job distribution to these sites is managed through \htc. 

Jobs running on external resources must be independent from each other and can access local {\submit} resources such as stored data only via \XROOTD, which is relatively slow. This restricts the scope of applications and makes certain tasks inefficient---such as those that are I/O limited---or even infeasible, such as those where jobs depend on each other.

To fill this gap a batch system providing access to local resources on {\submit} (2500 cores) is available through the {\slurm} job management tool. Applications like \DaskGateway~\cite{daskgateway} and OpenMPI~\cite{openmpi} are supported, which enable parallel communication and coordination between nodes in distributed systems. These tools streamline the process by transparently managing distributed computations through APIs within the user application. The main benefit of these tools is their ability to optimize the use of resources when executing computing workloads, freeing users from the burden of installing and coordinating the underlying infrastructure.

\subsection{Software availability}
With the growing effort in the computing community to develop more sophisticated software to advance scientific analysis, an increasing percentage of critical operational and management functions are enabled at the software layer rather than by the underlying hardware. 
The {\submit} system operates on the AlmaLinux\,9 operating system. However, many analyses require specific libraries tailored to their unique needs, making it challenging to run them within a single, generic software environment. Therefore, it is imperative for any analysis facility to provide several options for users to install, develop, maintain, distribute, and scale software for their workflows.

A key tool to achieve this are containers, which users can interact with using Apptainer or Podman.
These allow users nearly complete control over their own computational environment, and can be used both on the login node and on the distributed resources, or when moving from one facility to another.
The use of containerized software also guarantees reproducible and maintainable workflows.
Conda~\cite{conda} is used as a lightweight alternative for package and environment management.

A set of centrally maintained packages and libraries are available via the CVMFS distributed file system, including several CERN-related repositories available via \verb|/cvmfs/unpacked.cern.ch|.
Further, a local repository \verb|/cvmfs/cvmfs.cmsaf.mit.edu| is maintained, where additional software or data can be added by the support team, which is made available across all {\submit} nodes, as well as the MIT T3 and T2.

\subsection{Maintenance and upgrades}
The \submit~system requires periodic updates to maintain security, performance, and compatibility with modern software. During operation in the last two years, {\submit} underwent two major upgrades with the Linux update from Centos\,7 to AlmaLinux\,9 and the migration of the file system from GlusterFS to CephFS. 

When planning for system maintenance, it is essential to implement a phased approach. By upgrading the system in stages and providing test partitions, users were able to validate their workflows on the new environment before full deployment, reducing the risk of disruption to critical research activities. Throughout this process, transparent communication with the {\submit} users kept stakeholders informed about upcoming changes and temporary access limitations. Users that encountered migration issues provided valuable feedback and received timely assistance from the {\submit} team. This combination of technical solutions and user support made was crucial to successfully transition a complex multi-user system with minimal disruption to research activities.

For backward compatibility, we temporarily maintained access to the previous environment while transitioning to the new system. This approach allowed users to gradually adapt their workflows rather than facing an abrupt cut-off point. We identified that while many software components (particularly Conda environments) were portable between systems, workflows involving compiled code often required recompilation due to differences in system libraries. Containerization proved invaluable for managing legacy dependencies, as we provided containers of the previous operating system for workflows that could not be migrated immediately. This solution enabled continuity for users with specific version requirements or complex dependencies.
\section{System Usage and User Support}
\label{user}

The {\submit} analysis facility supports a diverse and growing user community, providing essential computational resources for research. This section explores the demographics of the users, the monitoring tools, and the support mechanisms that ensure the effective operation of \submit.

\begin{figure}
    \centering
    \includegraphics[width=0.31\textwidth]{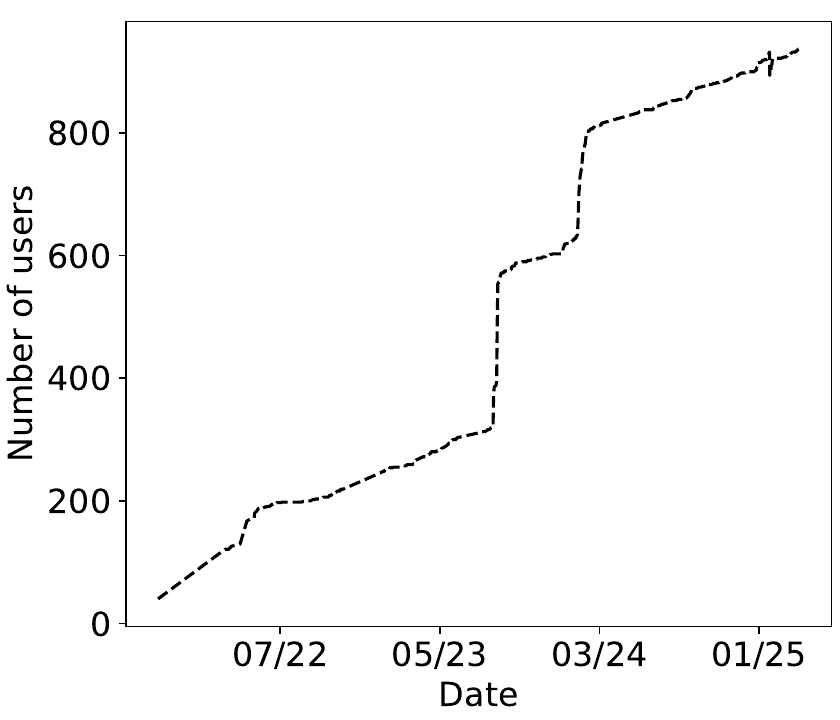}
    \includegraphics[width=0.665\textwidth]{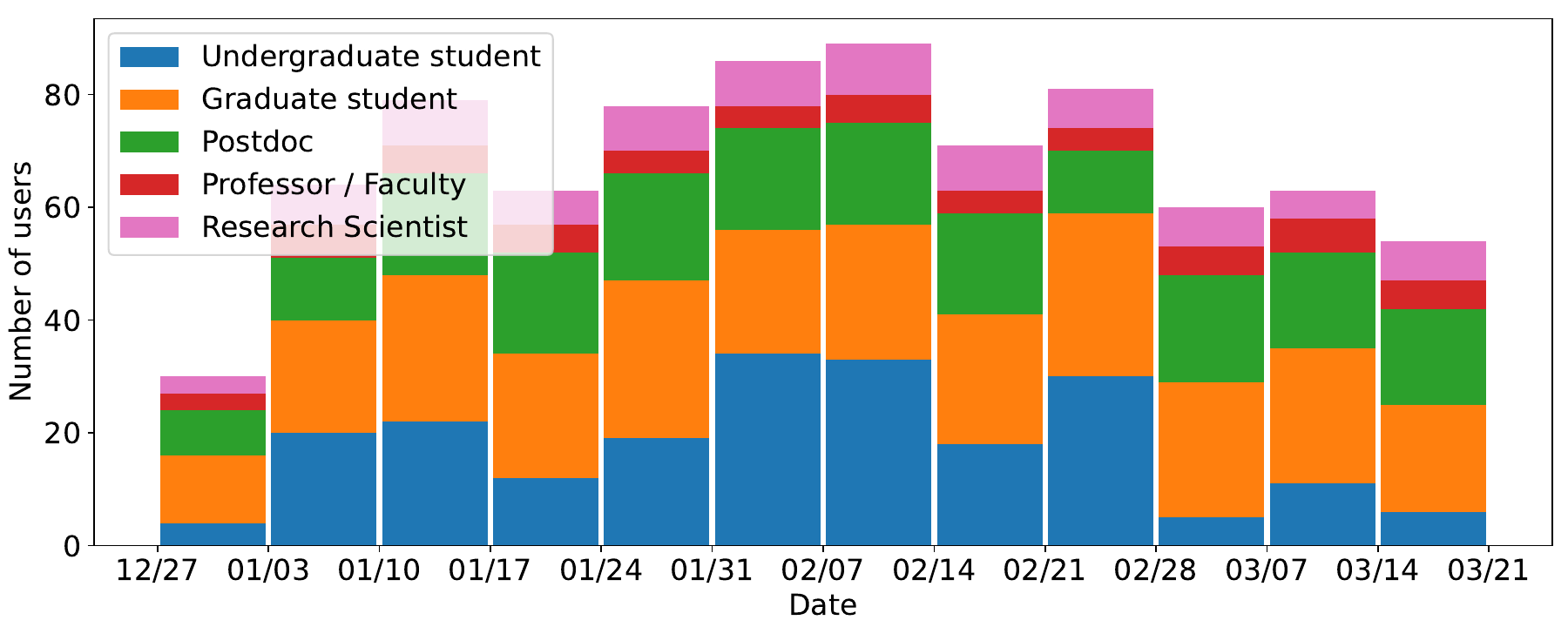}
    \caption{(left) Total number of registered users since opening the service to the department. (right) Number of distinct users that used {\submit} each week from December 2024 to March 2025. The users are categorized by academic position.}
    \label{fig:usersSubmit}
\end{figure}

\subsection{User survey}
Out of about 1000 registered users, about 800 use {\submit} for academic purposes, while approximately 200 use it for research---we will refer to them as research users---and access the platform regularly each month. The left panel of the Figure~\ref{fig:usersSubmit} shows all registered users over time, highlighting sharp increases in classroom usage in addition to a more shallow, steady increase due to research usage. The small reduction in January 2025 is due to the introduction of a user removal process for those that have left MIT, which will be performed on a yearly basis. The right panel of Figure~\ref{fig:usersSubmit} illustrates the weekly count of active users over the past few months, categorized by their academic positions.

\begin{figure}
    \centering    \includegraphics[width=0.6\textwidth]{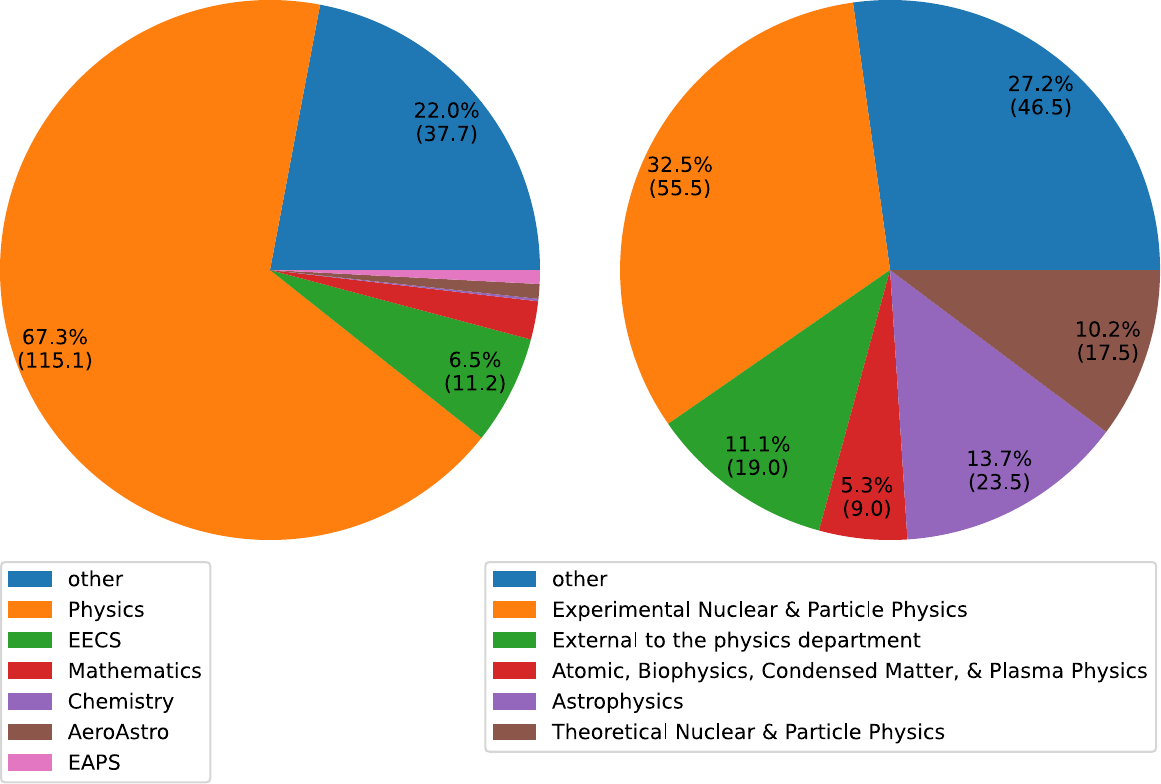} 
    \caption{Population of research users categorized by department (left) and division within the physics department (right). The category ``other'' also contains the users that are external to MIT. Indicated in brackets are the number of users per category. If a user is associated to multiple categories the user is counted in fractions to each contributing category.}
    \label{fig:survey1}
\end{figure}

\begin{figure}
    \centering
    \includegraphics[width=0.6\textwidth]{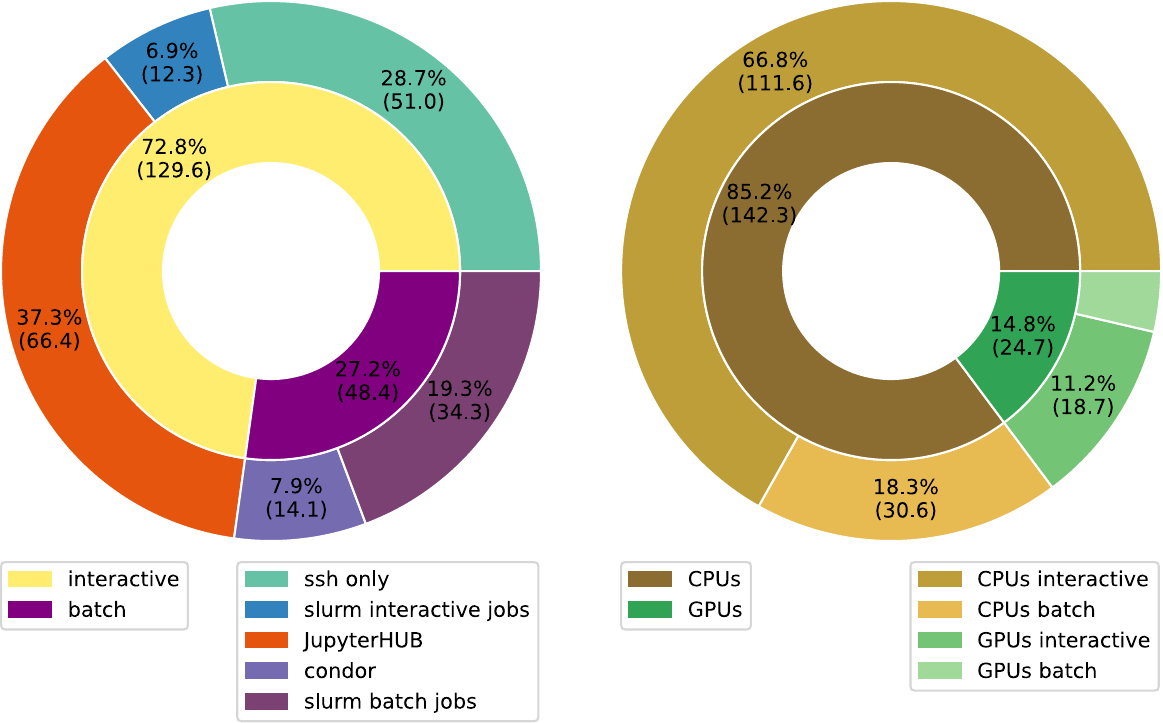}
    \caption{Type of user interaction with the system in terms of software (left) and hardware (right).}
    \label{fig:survey3}
\end{figure}

About 25\% of these users are external to MIT, which means that they are collaborating with MIT researchers. Figure~\ref{fig:survey1} illustrates the {\submit} research users population for active users from November 2024 to February 2025, providing a detailed breakdown of demographics and affiliations, showcasing the diversity and variety of backgrounds of the community. These pie charts focus on research users and therefore users for academic purposes, mostly undergraduate students, are not included. Figure~\ref{fig:survey3} displays how research users interact with the system. Approximately 27\% of research users submit batch jobs where about 70\% of them use internal resources through {\slurm} while others use external resources via \htc. 
Approximately 73\% of users engage with the system interactively. Among them, about half connect via SSH to login nodes, utilizing terminal sessions or allocating resources through {\slurm}. The remaining users access {\submit} through \jupyterhub, benefiting from a browser-based interface.
Also, 25\% of research users make use of GPUs, while 75\% rely on CPUs. The vast majority of academic users that are not included in Figure~\ref{fig:survey3} use the system interactively, via \jupyterhub, on CPUs.

Active users were asked in a survey, ``What are the key features that are essential for you on {\submit}?''. Users value different aspects of hardware resources, CPUs, GPUs, and storage, which validates our approach of a more heterogeneous structure of the system. For one in four users the support is an important component. Other users highlight the importance of resource and job management tools, as well as the availability of different software, such as Mathematica and Matlab. 

\subsection{User Support}
User support is provided through three primary methods. First, a comprehensive User's Guide~\cite{UserGuide} offers extensive and specific instructions. It includes a main section designed to help users get started, detailing all available resources (storage, software, batch systems, {\it etc.}). It also provides several examples of workflows and tutorials to facilitate onboarding. Second, a dedicated team of experts is available to answer user questions, accessible via email through a ticketing systems and a Slack help desk channel within the MIT Enterprise Slack workspace. Finally, automated assistance is provided by an AI-powered chatbot, AI Augmented Research Chat Intelligence (A2rchi).

This experimental application, A2rchi, uses a Large Language Model (LLM) for two purposes: it offers interactive user support through a chat interface, and it also enhances the efficiency of support ticket handling by generating draft responses for expert validation by the {\submit} team. A2rchi is trained on content from the User's Guide and internal {\submit} documentation, utilizing the Retrieval-Augmented Generation (RAG) method~\cite{RAG}, which combines custom data with general base knowledge. 

\subsection{Monitoring}
\begin{figure}[h]
    \centering
    \includegraphics[width=0.45\textwidth]{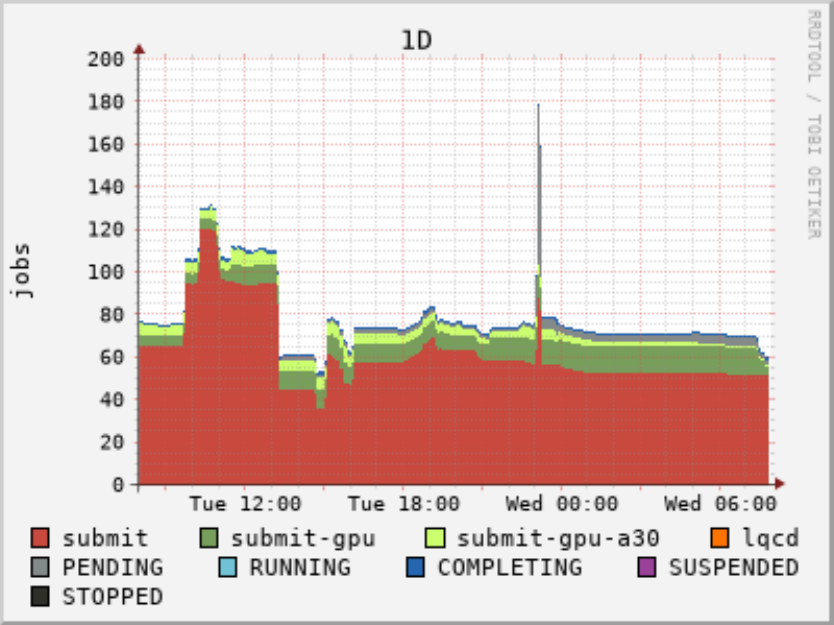}~
    \includegraphics[width=0.45\textwidth]{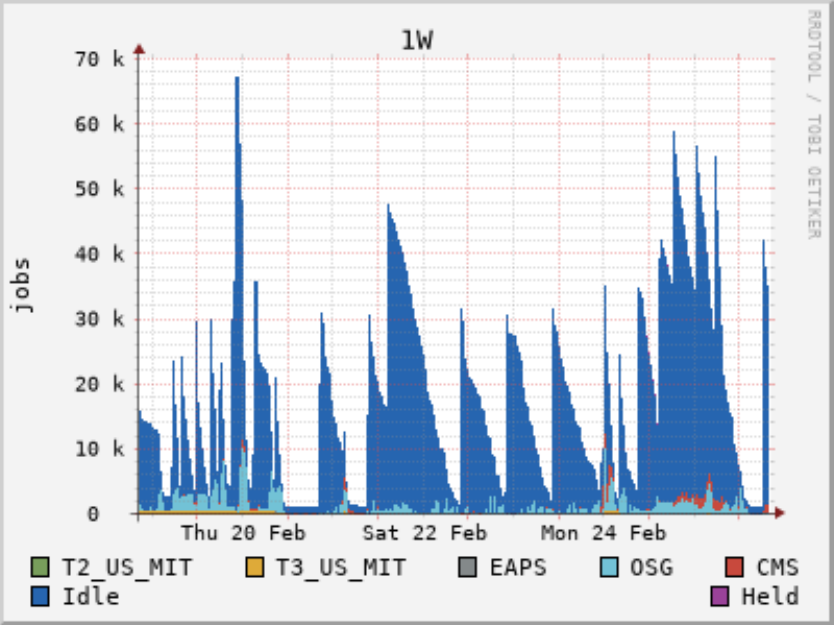}
    \caption{Example of monitoring images for the total number of jobs as a function of time submitted via {\slurm} for a time window of a day (left) or {\htc} for a time window of a week (right). The colors indicate the status of a job, and if running, on which partition or site it is being executed.}
    \label{fig:monitoring}
\end{figure}

A transparent public dashboard provides real-time insights performance and community metrics of \submit. Built on Round-Robin Database (RRD) tools~\cite{rrdtool}, this dashboard displays comprehensive information about batch job status and distribution across computing resources. Users can monitor both aggregate system usage and individual job statistics through interactive visualizations. The dashboard generates temporal plot summaries across multiple timescales (2 hours, 1 day, and 1 week), enabling trend analysis and resource utilization patterns. In addition, it offers detailed breakdowns of computing usage by {\htc} site and {\slurm} partition, allowing stakeholders to evaluate resource allocation efficiency and system performance at various levels of granularity. Figure~\ref{fig:monitoring} provides an exemplary overview of the total jobs submitted to {\slurm} and \htc.

Separately, Ganglia~\cite{ganglia} monitors all server machines within the {\submit} system, tracking metrics such as CPU usage, network activity, and memory. This monitoring is crucial for assessing the load on individual servers and ensuring efficient resource allocation.
\section{Analysis examples}
\label{analysis}
This section provides a comprehensive evaluation of the {\submit} system's performance.
Subsection~\ref{AGC} presents a benchmark assessment offering insights into its computational capabilities and efficiency.
Subsection~\ref{cmsanalysis} explores concrete examples from the CMS experiment, illustrating practical applications and the {\submit} system's performance in real-world analysis scenarios.
Subsection~\ref{FCC} reports analysis workflows related to the FCC study, highlighting the adaptability of the {\submit} system to different experimental frameworks.
Beyond research applications, {\submit} plays a pivotal role in educational initiatives. 
Subsection~\ref{edu} discusses how the system is utilized in academic settings, facilitating hands-on learning experiences for students and educators alike.
Collectively, these diverse aspects address varying levels of expertise, ensuring accessibility for all users.

\subsection{Analysis Grand Challenge}
\label{AGC}
The Analysis Grand Challenge (AGC)~\cite{AGC,Held:2022sfw} project establishes a benchmark analysis using open data from the CMS experiment~\cite{OpenData}, originally collected in 2015. This serves as a realistic workflow that can be used to compare different frameworks, provide feedback to developers of a specific framework, and validate different aspects of an analysis facility. The AGC task focuses on a top quark pair production measurement and includes event selection, weighting, definitions of new observables, application of systematic variation, ML inference, and aggregation into histograms for plotting. In this Section, we report on experimental measurements of the analysis throughput.

The input dataset total size is 1.7\,TB, of which only roughly 5\% is read by the analysis task, which is typical for such studies. The software environment is based on {\ROOT} (version 6.32.08), using {\RDF} as the high-level interface for the data analysis. While the high-level interface is separately optimized for this exercise~\cite{AGCROOT}, the AGC analysis is used to validate the key system aspects of the {\submit} system such as I/O performance and distributed computing, by comparing different configurations.

\subsubsection{Test of the I/O}
To evaluate the input read performance, tests were conducted on a high-performance machine equipped with 192 CPU cores and 1.5\,TB of memory, ensuring that the I/O subsystem was the only limiting factor. The {\nano} dataset was replicated on various systems, including \verb|/scratch/submit| ({\NVME} disks), \verb|/ceph/submit| (CephFS distributed file system on spinning disks), MIT \Ttwo~or accessed remotely through either an Xcache server (\verb|root://bost-cms-xcache01.es.net/|) or directly via {\XROOTD} from CERN-EOS and FNAL-EOS.

\begin{figure}[h]
    \centering
    \includegraphics[height=0.27\linewidth]{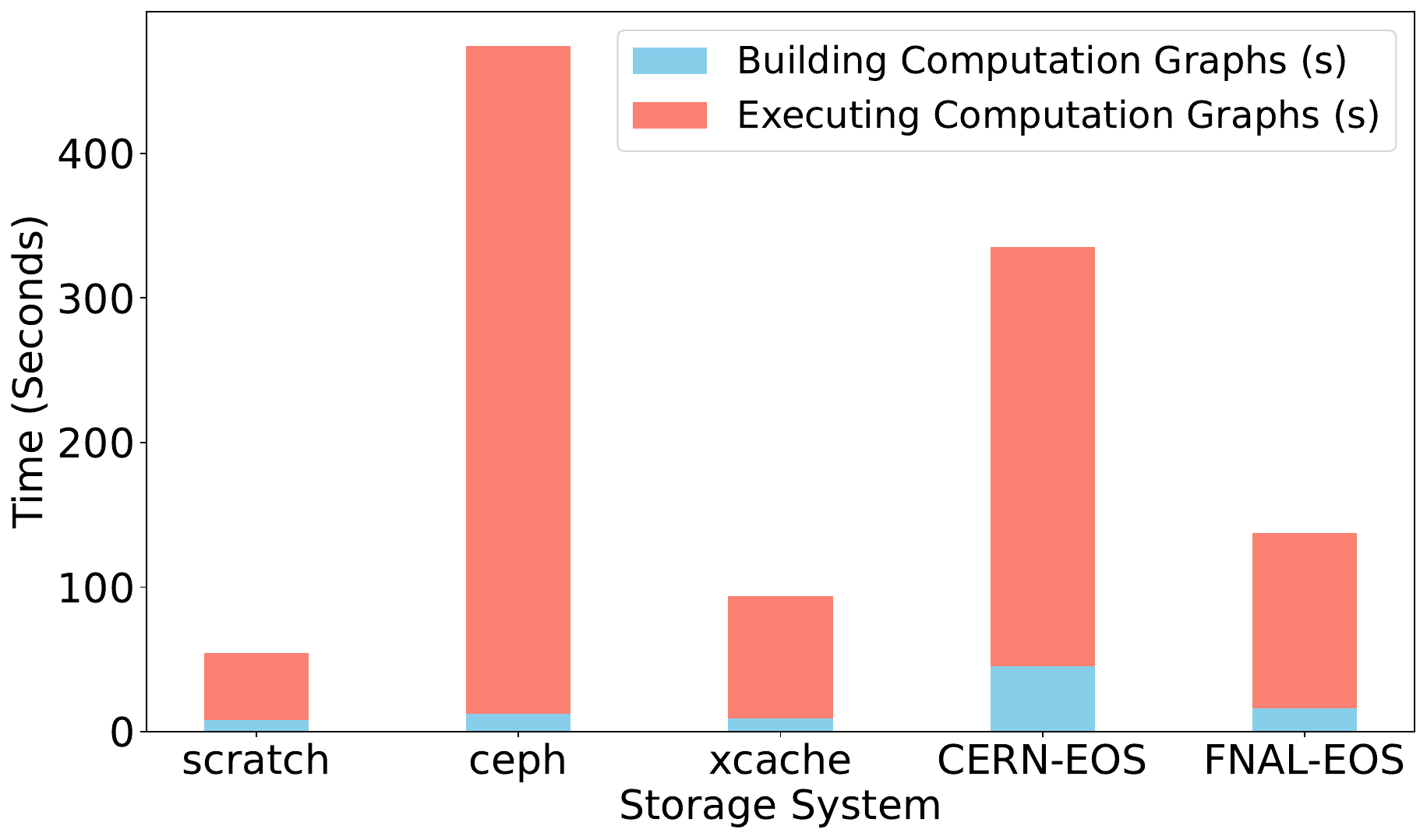}
    \includegraphics[height=0.27\linewidth]{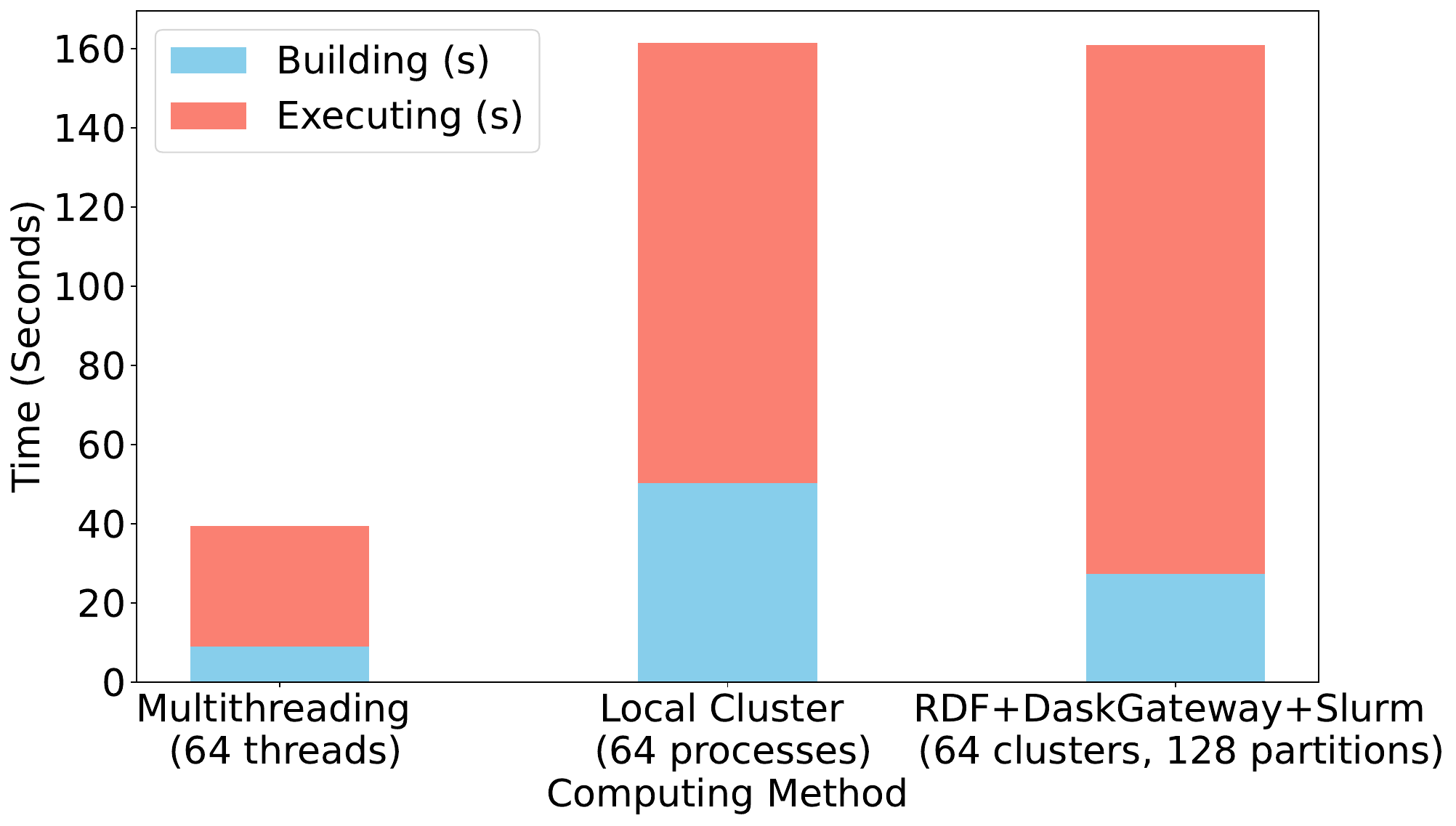}
    \caption{Left: Performance comparison across storage system: time spent to read from different input sources. Right: Performance comparison across computing configurations: time spent to execute the task locally on a node or remote on multi-nodes.}
    \label{fig:AGC}
\end{figure}

The results, shown on the left side of Figure~\ref{fig:AGC}, confirm the fastest data reads were achieved using the {\NVME} disks connected with the {\centoGbps} link, as expected. The Xcache service, also utilizing {\NVME} disks, offers a fast read though it remains slower than local disk reads. Both FNAL-EOS and CERN-EOS demonstrated efficient performance even reading from remote. 
The CephFS system exhibited slower performance due to its buffered I/O mechanism, which enforces a minimum read-ahead size of 16\,MB in our current configuration, significantly larger than the typical data required per read for this application. The MIT {\Ttwo} access, which features a comparable CephFS-based storage system and network link speeds, showed performance levels comparable to \verb|/ceph/submit|.

The I/O performance is also expected to improve with the {\ROOT}'s team development of the direct I/O and RNTuples~\cite{RNtuples}, a columnar storage format for event data optimized for selective reads. 

\subsubsection{Distributed execution}
The distributed {\RDF} application transparently runs on one or multiple machines. On the right side of Figure~\ref{fig:AGC} we compare the native parallelization on a single high-performance machine using implicit multithreading and the distributed execution across multiple nodes. In the distributed setup, {\RDF} connects to the {\DaskGateway} Server, which interacts with {\slurm} to allocate computational resources. To eliminate I/O latency as a limiting factor, data is read from a local {\NVME} disk over a {\centoGbps} link.

The results indicate that, as expected, the preparation and the retrieval of the distributed tasks introduce some overhead compared to intrinsic multithreading on a single node. However, we observe no additional scheduling overhead when allocating resources across multiple nodes, demonstrating the efficiency of the \DaskGateway-{\slurm} integration for distributed computing.

\subsection{CMS Analyses examples}
\label{cmsanalysis}

The CMS analyses presented in this section leverage {\submit} as the primary user access point, demonstrating its flexibility in supporting diverse research workflows. These studies helped identify key bottlenecks in traditional analysis approaches and explore modern solutions to improve efficiency.

Historically, HEP data analysis rely on custom loops over {\ROOT} datasets using experiment-specific tools. Users write code independently of the system that handles job distribution. This approach separates user code development from job execution, making it difficult to maintain a consistent software environment across platforms. Researchers often spend more time on data preparation, job batching, and result aggregation than on the actual analysis. To address these inefficiencies, we adopted modern computing tools and streamlined workflows on {\submit}.

Key improvements include direct data access to minimize transfer delays by working with centrally managed storage. Improved data management practices avoid unnecessary duplication of datasets, ensuring that researchers can reuse existing data. Automated computation management handles event processing and distribution transparently. Advanced computing tools such as GPU acceleration, multithreading, and parallelized workflows enhance processing speed. Finally, simplified software deployment using containers improve reproducibility and ease of use.

This section highlights four exemplary CMS analyses conducted on {\submit}, covering the full workflow from initial design to final publication:
\begin{itemize}
    \item a Standard Model measurement of the opposite-sign W boson pair cross-section (WWos)~\cite{CMS:2024hey},
    \item a high-precision measurement of the W boson mass (MW)~\cite{CMS:2024lrd},
    \item searches for rare Higgs boson decays (Hrare)~\cite{CMS:2024tgj}, and
    \item an exotic search for Soft Unclustered Energy Patterns (SUEP)~\cite{CMS:2024nca}.
\end{itemize}
Table~\ref{tab:AnalysisFeatures} summarizes the tools used in each case. Each analysis approaches a typical staged workflow: data preparation followed by data processing and histogram creation, and finally statistical interpretation and visualization. 

\begin{table}[]
    \centering
    \caption{CMS data analyses executed on {\submit} and the specific features used.}
    \label{tab:AnalysisFeatures}
    \begin{tabular}{l|c|c|c|c}
               & WWos            & Hrare         & SUEP         & MW \\
\hline
data volume    & 7.5\,TB          & 50\,TB         & 100\,TB       & 16\,TB\\        
input format   & \nano           & custom   & custom  & custom \\
               &                 & \nano    & \nano   & \nano \\ 
\hline
environment    & CMSSW         & conda           & conda & Singularity \\
interface      & \RDF          & \RDF            & \coffea   &  \RDF  \\
compute        & \slurm/\htc   & multithreading  & \slurm & multithreading\\
storage        & /ceph         & /ceph \& /scratch & /ceph & /scratch\\
MVA            & TMVA          & TMVA-XGBoost    &  &  \\  
\hline
histogramming  & \ROOT         & \ROOT           &  ROOT \& Boost       & Boost \\
plotting       & Python shell  & Python shell    & Jupyter & Python shell \\
stat. analysis & Combine       & Combine         & Combine & CombineTF\\ 
    \end{tabular}
    
\end{table}

\subsubsection{Data preparation}
All analyses described here are based on the {\nano} data format, which allows decoupling of the core analysis process from the complex CMSSW framework. For the WWos analysis, which is the most standard CMS analysis of the four, the official {\nano} sample is used as input, because no modifications are required for this measurement. The other three analyses require more detailed information, such as reconstructed tracks or particle flow collections~\cite{PF}, detailed detector data for calibration, and extended theory inputs. This is not available in the official {\nano} samples and custom {\nano} samples are produced and stored separately.

These additional steps require effort, CMSSW expertise, computing power, data storage space, and is part of the analysis process. The modifications to the standard {\nano} workflow are minimized by taking advantage of dedicated libraries within CMSSW, such as kinematic fitting of vertices, without the need to re-implement these features from scratch. These customized {\nano} files are saved without any event selection, enabling their use in other analyses with similar needs. This processing step relies on the WLCG computing and storage resources. These custom {\nano} files are made available to the analysis collaborators via \XROOTD. To conserve resources, the central {\nano} version undergoes continuous optimization, aiming to support the maximum number of analyses while maintaining a manageable dataset size.

During the analysis design phase, simulated samples are often privately generated and processed with the full reconstruction sequence. A better practice developed within the MW analysis is the direct production of so-called NanoGEN files that follows the {\nano} format but without detector simulation. This allows for fast validation and optimization before being sent to central production where larger numbers of events can be generated. The grid global pool provides the necessary computing power, while the {\nano} files for both simulation and data are stored at the MIT Tier-2 accounting for approximately 10-100\,TB storage.

\subsubsection{Data processing}
A fast turnover in the event loop is crucial for enabling continuous analysis development at a rapid pace. The {\RDF} environment, which has evolved from the {\ROOT} framework, is a common choice for managing data structures in HEP (e.g., WWos, Hrare, MW), while {\coffea} serves as an alternative and more enterprise oriented eco system (e.g., SUEP).

The MW analysis has a high selection efficiency and uses a large fraction of the stored information. Thus, the full custom {\nano} is processed in one step. 
Other analyses (WWos, Hrare, SUEP) begin with data preparation, including skimming and slimming. Skimming is typically based on some given trigger paths and final state configuration. The next step is the main processing loop, which is executed interactively to profit from the {\RDF} multithreading (Hrare, MW) or from the batch via {\slurm} or {\htc} (WWos, SUEP). This phase involves event selection, construction of observables, application of corrections, and evaluation of systematic variations. At this stage, files are stored in convenient disk systems, such as local spinning disk or {\NVME} based file systems if possible. These are used for both histograms and Multi Variate Analysis (MVA) training. The MVA training results are then applied during the next re-run of the analysis. Variations to determine systematic uncertainties are applied on the fly to minimize the volume of stored data. \corrLib~\cite{correctionLib} is a CMS centralized library used to apply simulation-to-data scale factor corrections and handle systematic variations. 

The result of the data processing is output in histograms and stored on disk. Traditionally, ROOT histograms are stored in ROOT files (WWos, Hrare, SUEP).
In the MW analysis the nominal histograms covers muon kinematics and properties for in-situ background estimation with around 20,000 bins. Histograms for systematic variations get as large as 20,000,000 bins. This required the integration of Boost histograms~\cite{BoostHist} with atomic aggregation in C++, including custom axis types providing higher memory efficiency and flexibility. The objects are then pickled and stored in HDF5 files to allow for fast on-demand access.

To ensure stability and reliability, the software environment is managed using Conda (Hrare), custom Singularity containers (MW, SUEP), or the version provided within CMSSW (WWos).

\subsubsection{Visualization and statistical interpretation}
The final stage of these analyses involves producing figures for visual inspection of the results and the quantitative assessment of physical observables in a statistical interpretation. Plotting is handled directly by ROOT or python, either in scripts executed interactively or in jupyter notebooks. The possibility of having a personal webpage on subMIT to inspect and share them with collaborators comes in handy. 
Statistical interpretations are traditionally performed using the CMS Combine tool which is based on RooFit~\cite{CMS:2024onh}. 
For more complex problems, such as for the MW analysis, the CombineTF framework was developed to exploit automatic differentiation through the TensorFlow library~\cite{combinetf2}. In combination with state-of-the art minimizes and multi-threading capabilities, this leads to a vast acceleration of this step.

\begin{figure}[h]
    \centering
    \includegraphics[width=0.75\textwidth]{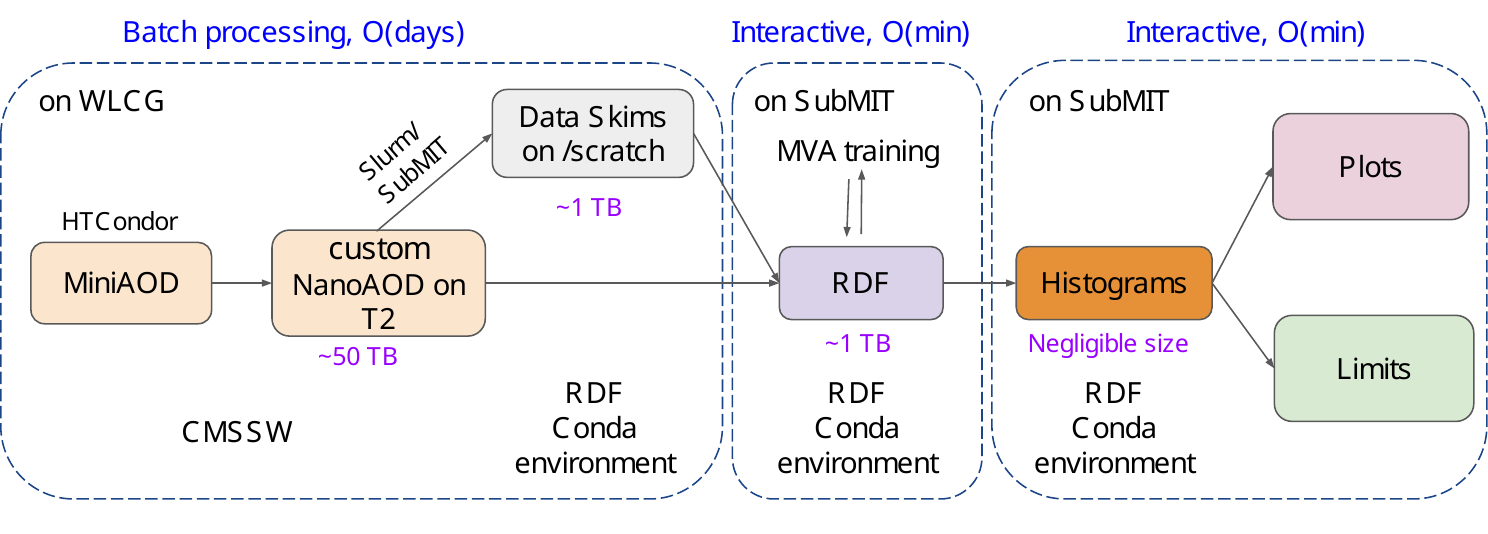}\\
    \caption{The Hrare analysis workflow. The computing resources that are used to analyse the data are shown above each step while the approximate storage space is shown below in purple. The dashed boxes represent different stages of the analysis that require different software as shown in the bottom. The approximate time required to process files for each stage is shown at the top in blue.}
    \label{fig:HrareFlow}
\end{figure}

Figure~\ref{fig:HrareFlow} shows the workflow of Hare analysis performed through the {\submit} system. The Data processing step was also replicated at the Purdue AF and the Fermilab EAF with the custom {\nano} accessed from the MIT \Ttwo.

\subsection{FCC experience}
\label{FCC}

The FCC is a proposed accelerator at CERN, initially planned to operate as an electron-positron collider for 15 years (FCC-ee) before transitioning to a proton-proton collider (FCC-hh), similar to the LEP-LHC tandem. A feasibility study is currently underway to evaluate the physics potential, detector requirements, and operational conditions of the accelerator. These assessments are being conducted using simulations and modern analysis tools derived from those employed for research at the LHC.

A variety of analyses are being conducted at MIT to support the FCC feasibility study, with a focus on FCC-ee, including:
\begin{itemize}
    \item measurements of the Higgs boson mass, cross-section, and couplings,
    \item measurements of the W boson mass, using the threshold scan technique as well as with kinematic fits,
    \item measurements of the Z boson mass, width, and cross sections at different center-of-mass energies, and detailed studies of backgrounds in those analyses,
    \item studies of detector designs based on parametric and full hit based simulations, and
    \item studies of beam background.
\end{itemize}
In addition to supporting the feasibility study, the FCC framework serves as an educational platform for both undergraduate and graduate students. Students typically do not have a CERN account but they can get one for {\submit}. It allows for simple analyses without requiring detailed background knowledge in physics or programming. Several examples are made available for students to perform reasonably complete analyses of various kinds. The details of each workflow depend on the individual analysis, but a global summary is given below.

Central to each workflow is Key4hep~\cite{key4hep}, a software framework designed to support the development and operation of experiments in HEP. It aims to provide a unified, common set of tools and interfaces that is used across different experiments. Key4hep is available from CVMFS and can be loaded by sourcing a setup script.

As the studies are based on simulation, event generators are crucial for producing the necessary physics processes. While Key4hep offers a variety of generators, it is extended with state-of-the-art generators. Most of the production is done central by the FCC community, but private campaigns to use additional generators have been produced using {\submit} via \htc. After being produced, the events are processed using Delphes~\cite{Delphes} for a parametric detector simulation, probing different detector designs. The final objects are stored in {\ROOT} files in the EDM4hep data model format, which is integral to the Key4hep framework. A total of 20\,TB of generated events are being actively used, stored on the \verb|/ceph/submit| disk.

The analysis is based on the {\RDF} framework to efficiently loop over the events. The necessary tools to read the EDM4hep data format are available in Key4hep, which also offers many additional functionalities such as jet clustering, flavor tagging, and physics utilities specific to electron-positron colliders. The integration of these tools in the central Key4hep repository makes it straightforward for students to work with and interact with these tools.

Typically, the analyses are performed interactively, directly producing histogram {\ROOT} files by looping over the events in the EDM4hep files. The interactive use is intuitive for starting students, making it easier for them to engage with the analysis process. Additional options are available to reduce or skim the events or to add more predefined content if desired, particularly for jet clustering or flavor tagging, as these operations are typically slow. Functionality is also available to scale out the processing to batch computing systems {\htc} and {\slurm}, providing flexibility. 

The final products of the analysers are histograms, which are further processed through plotting or fitting. Typically, this last step of the analysis workflow occurs using Jupyter Notebooks hosted on the {\jupyterhub} instance, where the histograms can be read, processed, fitted, and plotted using several Python packages which are popular in the HEP community, such as uproot, matplotlib, mplhep, and scipy. For undergraduate students having their first experience with a HEP analysis, Python environments containing these packages are centrally created with Anaconda, and distributed through CVMFS, allowing the students to focus on the physics goals and reduce the computational expertise needed for entry in the field. More systematic plotting scripts can be used outside Jupyter Notebooks, and their outputs are displayed on the local web area available on {\submit} to each user. Statistical analysis is typically performed using the CMS Combine tool or CombineTF. Examples and documentation are provided for students to set up the necessary input files to run the statistical analysis and interpretation.

An example of such a workflow is the analysis of the Higgs boson mass. This analysis requires a tight event selection in order to reject the large backgrounds. The outcome of the event loop in {\RDF} are histograms of the recoil distribution, which are analytically parameterized using RooFit, and then injected into Combine to run a statistical fit in order to extract the uncertainty on the measurement of the Higgs boson mass. A total of 10\,TB of samples containing about 100 million events, are typically read interactively and the process can be completed in 30 minutes by reading from \ceph. Typically the reading speeds of {\ceph} processing saturates the available network bandwidth of {\centoGbps} and the number of cores used to run the {\RDF} framework can be limited to 32 or 64 to optimize CPU usage for other processes. This underlines the importance of a well balanced systems were networking is matched with the storage performance.

\subsection{Education}
\label{edu}
While the primary aim of {\submit} is research, there has been a growing interest for the use of {\submit} in classroom settings. Several undergraduate and graduate courses at MIT have benefited from the computing resources available on \submit, and have worked with the {\submit} team to help define standard procedures for student interactions with the cluster. These include introductory undergraduate physics courses in classical mechanics, electromagnetism, and electrostatics,
as well as advanced undergraduate and graduate courses. A knowledgeable and available {\submit} team that can be relied on by instructors to implement their ideas has emerged as an important cornerstone for effective use of the cluster resources.

\subsubsection{Student access}
Student access for education is identical to any other user accessing the cluster. Since every MIT student already has an MIT ID, access to the cluster needs only approval from the {\submit} team. Students cross-registered from other universities obtain an MIT ID early in the semester. The instructors coordinate the {\submit} approval prior to the start of the semester by providing a list of student names or IDs, which are then registered as {\submit} users in bulk, avoiding tedious exchange of emails.

\subsubsection{Software distribution}
Courses usually require specific software that the students would need to access easily and concurrently. These can range from simple Python environments to specialized programs used in research.
The instructors send their software, or recipes for installing it, to the {\submit} team which then places it on the CVMFS distributed file system. This makes the system robust even when hundreds of students are accessing the software at the same time, a common occurrence in a large classroom during lecture, recitation (a period where students are expected to do hands-on work), or prior to assignment deadlines. The software can then be accessed from anywhere on the cluster by any user. In the case of a Python environment, a globally-visible kernel can also be created, which appears as an icon for all students on the {\jupyterhub} landing page after login. By clicking the icon, as with privately-created kernels, a notebook is launched which uses the desired python environment. 
Furthermore, when a student opens a notebook prepared by an instructor using one of these custom environments, the appropriate kernel is automatically loaded. 
This streamlines access to custom environments, requiring no extra steps from the students.
Some environments required additional server-side Jupyter extensions, which were installed by the {\submit} staff, becoming available to all users.

\subsubsection{Resource utilization}
During times of particular interest, such as during lectures or in the days before assignments are due, it is necessary for students to have some guaranteed resources that they can access without delay.
This is ensured via {\slurm} reservations, which can block out an arbitrary amount of nodes or CPUs, for a particular group of users, and for particular hours of any day of the week. Since {\jupyterhub} is configured to launch servers also using \slurm, an option in the spawn page is added to use the appropriate {\slurm} reservation.

\subsubsection{Example 1: Introductory courses}
Introductory undergraduate courses at MIT are characterized by the Technology-Enabled Active Learning (TEAL) format~\cite{Dourmashkin2020}, in which classroom time is spent working in small groups to solve exercises and conduct small experiments, often with the aid of computers. Being required courses for all undergraduate students, class sizes are in the O(100) with several sections per week, composed of many students who are not pursuing physics as their primary area of study. Here {\submit} is used as a support infrastructure providing LLM access that has been enhanced with course specific information using a RAG mechanism. The platform also allows students to run simulations that visualize electric and magnetic fields.

\subsubsection{Example 2: Advanced courses}
Advanced undergraduate and graduate courses typically have smaller class sizes than introductory courses, but may make more intense use of the resources and have more specialized software needs. It is common for these courses to rely on the {\submit} team to help install their specialized software on \submit, leveraging the team's experience with scientific software and the cluster architecture and operating system.

For example, an undergraduate course in modern astrophysics used the cluster to provide computing resources and a consistent environment for students to carry out assignments through \jupyterhub. Astrophysics software such as AstroPy~\cite{astropy} was installed into a Conda environment, and distributed through CVMFS. In another case, an advanced graduate course on astrophysics and cosmology used the cluster to run and analyse cosmological simulations with the Gadget~\cite{gadget} software package, which relies on OpenMPI, through \slurm.

Another set of classes that make use of {\submit} are the undergraduate physics laboratory classes at MIT (8.13 and 8.14), where the students perform various types of data analysis, including some CMS Open Data based analyses, which {\submit} offers strong support for.

The main point of {\submit} is that it comes with a rich and flexible environment and that there are detailed instructions for helping the user with the most commonly used software packages. For example people using for standard python or {\ROOT} tools will be able to perform data analysis without having to deal with installing software on their own laptop without dedicated support and there is a well defined way how to install additional software.

\section{Conclusion}
\label{conclusion}

In summary, the {\submit} platform represents a comprehensive and robust prototype of an Analysis Facility that meets the demands of the HL-LHC era, but also successfully serves a variety of other communities. We have successfully performed a number of real-life LHC analyses from beginning to publication and proven that {\submit} empowers users to conduct data analyses interactively, achieving turnaround times on the order of minutes even when working with terabyte-scale datasets. Medium-scale data processing tasks can be executed seamlessly by offloading jobs to local batch resources, while large-scale workflows benefit from tight integration with distributed computing systems such as the WLCG and the OSG. {\submit} also offers advanced capabilities for managing and accessing large datasets, streamlining the analysis process for a broad spectrum of users.

The success of the platform stems from several crucial characteristics. These include that each server is equipped with at least one {\centoGbps} network connection, ensuring rapid data access and communication; high-performance hardware featuring maximum core count servers with generous memory allocations; and dedicated GPU resources for machine learning and other workloads that benefit from parallel processing. Other essential characteristics are the storage infrastructure, which is highly optimized, combining expansive ultrafast scratch space ({\NVME}) and large-scale high-performance CephFS spinning disk storage.
User access is straightforward and secure, supported by two-factor authentication and users can manage their own software needs relying on the flexible software environment managed via containers and package managers. Last but not least, {\submit} offers robust user support, combining an engaged support team with a specialized AI-driven chatbot to help users navigate and utilize the facility effectively.

Together, these features make {\submit} a powerful, scalable, and user-friendly platform that serves not only as a model for future analysis facilities but also as a practical, working system that is already enabling high-impact scientific research and education.

\bibliography{sn-bibliography}

\end{document}